\documentclass[pra,twocolumn,showpacs]{revtex4}
\usepackage{graphicx}
\usepackage[usenames]{color}

\newcommand{\beq}{\begin{equation}}
\newcommand{\eeq}{\end{equation}}
\newcommand{\beqa}{\begin{eqnarray}}
\newcommand{\eeqa}{\end{eqnarray}}
\newcommand{\ba}{\begin{array}}
\newcommand{\ea}{\end{array}}

\begin{document}

\title{Nonlinear Schr\"odinger equation for a superfluid Bose gas \\
from weak coupling to unitarity: Study of vortices} 
\author{S. K. Adhikari$^{1}$\footnote{adhikari@ift.unesp.br; 
URL: www.ift.unesp.br/users/adhikari} 
and 
L. Salasnich$^{2}$\footnote{salasnich@pd.infn.it; 
URL: www.padova.infm.it/salasnich}}
\affiliation{
$^1$Instituto de F\'{\i}sica Te\'orica, UNESP - S\~ao Paulo State
University, 01.405-900 S\~ao Paulo, S\~ao Paulo, Brazil \\
$^2$CNR-INFM and CNISM, Unit\`a di Padova, 
Dipartimento di Fisica ``Galileo Galilei'', Universit\`a di Padova, Via
Marzolo 8, 35131 Padova, Italy}

\begin{abstract}
We introduce a nonlinear Schr\"odinger equation to describe the dynamics 
of a superfluid Bose gas in the crossover from the weak-coupling 
regime, where $a n^{1/3}\ll 1$ with $a$ the inter-atomic 
s-wave scattering length and $n$ the bosonic density, to 
the unitarity limit, where $a\to +\infty$. We call this equation 
the {\it unitarity Schr\"odinger equation} (USE). 
The zero-temperature bulk equation of state of this USE 
is parametrized by the Lee-Yang-Huang low-density expansion and 
Jastrow calculations at unitarity. With the help of the  USE 
we study the profiles of quantized vortices and vortex-core radius 
in a uniform Bose gas. We also consider quantized vortices 
in a Bose gas under cylindrically-symmetric harmonic 
confinement and study their profile and chemical potential using the USE 
and compare the results with those obtained from the 
Gross-Pitaevskii-type equations valid in the weak-coupling limit. 
{Finally, the USE is applied to calculate the breathing 
modes of the confined Bose gas as a function of the scattering length.}

\end{abstract}

\maketitle

\section{Introduction} 

For the theoretical investigation of the Bose-Einstein 
condensate (BEC) 
of an ultra-cold bosonic gas the main tool is 
the mean-field Gross-Pitaevskii equation (GPE) 
\cite{gross,rmp}, that is reliable for small values of the gas parameter 
$a n^{1/3}$, where $a$ is the s-wave scattering length of the 
inter-atomic potential and $n$ is the bosonic density. 
By manipulating a background magnetic field near  a Feshbach resonance 
the s-wave 
scattering length $a$ can be modified 
and can reach very large values corresponding to a strong  
atomic interaction \cite{cornish}. 
To take into account the effect of a large scattering length, 
a modified GPE (MGPE) has been introduced by Fabrocini 
and Polls \cite{polls1} by using the first two terms in the 
Lee-Yang-Huang expansion \cite{huang} 
of the energy of a uniform Bose gas, which include terms of the order 
of $\sqrt{na^3}$. 
Within the density functional approach of Fabrocini
and Polls \cite{polls1}, the leading term of expansion 
gives the GPE while the next term is responsible for
corrections due to  moderate atomic interaction. These two leading terms 
have been used to calculate beyond-mean-field corrections to properties 
of trapped BECs \cite{str}. The contribution terms of the order of 
$na^3$ has also been discussed in the literature \cite{not}.

In this paper we generalize the MGPE \cite{polls1}
by considering also 
the behavior of the bosonic system in the unitarity limit, where 
$a\to+\infty$. Jastrow calculations \cite{cowell,heis} suggest that 
in the unitarity limit the zero-temperature 
bulk chemical potential $\mu$ of the Bose system is 
given by $\mu = \xi n^{2/3}{\hbar^2/m} $, 
where $\xi$ is a constant and $m$ is the mass of a single atom. (Bulk 
chemical potential is essentially the nonlinear term that appear in the 
mean-field equation and is related to the energy per particle of the 
system.) 
This 
result is independent of the atomic scattering length.
In the 
present work the equation
of state of the bulk system is parametrized with a Pad\`e 
approximant by using the first two terms of 
the Lee-Yang-Huang expansion \cite{huang} and the Jastrow calculations 
at unitarity \cite{cowell,heis}. In this way we obtain 
a time-dependent highly-nonlinear 
Schr\"odinger equation, that we call unitarity 
Schr\"odinger equation (USE). The USE in general form is time dependent 
and can be used to study non-stationary dynamics, whereas its 
time-independent form is appropriate to study stationary states. 
The USE gives the hydrodynamic equations of bosonic superfluids 
at zero temperature, and enables one to study collective
dynamical 
properties of the system in the full crossover 
from weak-coupling to unitarity. 

As an application of the USE, 
here we study the structure of quantized vortices in both 
uniform Bose gas and Bose gas under axially-symmetric harmonic 
confinement 
and compare and contrast the results with those obtained with the MGPE
\cite{polls1} and GPE.
Quantized vortices in superfluids 
are a manifestation of quantum mechanics 
at the macroscopic level \cite{annett}. 
Recently quantized vortices have been observed 
in rotating ultra-cold atomic BECs \cite{raman} 
and also in atomic Fermi gases in the BCS-BEC crossover 
near a Feshbach resonance \cite{zwierlein}.  
Using the GPE, 
quantized vortices in a harmonic trap have been analyzed 
by Dalfovo and Stringari \cite{franco-vor}. More recently 
vortices have been investigated with GPE in various problems; 
for instance, collective modes of a vortex \cite{fetter-vor}, 
vortex under toroidal confinement \cite{rokhsar,sala-vor}, 
vortex lattice \cite{ueda}, 
vortex with attractive scattering length \cite{sadhan-vor,sala-vor1}, 
collision dynamics of vortices \cite{ska}, collapse of a vortex state 
\cite{ska2}, free expansion of vortices \cite{ska3}, 
and vortex in Bose-Fermi mixtures \cite{sala-vor2}. 
Nilsen {\it et al.} \cite{polls3} used the MGPE \cite{polls1}
to investigate the structure of vortices, and found 
a good agreement 
 between the MGPE and variational Monte Carlo 
results. Using the USE,
we extend the study  of  Nilsen {\it et al.} \cite{polls3} 
and find  that the radius of the vortex core 
decreases with the 
increase of  the scattering length. At the unitarity limit it reaches 
a critical minimal radius. The properties of this 
minimal radius depend on the trapping geometry: in the case of a vortex 
in a uniform Bose gas it is a decreasing function of the 
uniform density at large distances; in the case of a vortex 
in the harmonic trap it is a decreasing function of the total 
number of atoms. 
{Using the USE, we also calculate the 
radial and axial frequencies 
  of collective oscillations from weak-coupling to unitarity 
in a cigar-shaped trap.}

In Sec. \ref{II} we introduce the present model and relate it to 
superfluid hydrodynamics. Formulation for quantized vortices in the 
present  model is considered in Sec. \ref{III}. Section  \ref{IV} is 
devoted 
to vortex in a uniform Bose gas, where we numerically study the vortex 
profile and 
vortex-core radius for different scattering lengths. In this case the 
USE is solved by the fourth-order Runge-Kutta method \cite{RK,num}.
The vortex-core 
radius decreases with increasing scattering length and saturates to a 
constant value in the unitarity limit.  
In Sec. \ref{V} we consider a vortex in a Bose gas under 
axially-symmetric 
harmonic pancake-shaped confinement by solving the USE by imaginary time 
propagation using the semi-implicit Crank-Nicholson rule 
\cite{num,sala-numerics,cn}. 
In this case 
we study the profiles of vortices and corresponding chemical potentials 
and compare the results with those obtained from the MGPE 
\cite{polls1} and GPE.
The interesting feature of the results of the USE is that they 
saturate in the unitarity limit as $a\to \infty$, whereas the results of 
MGPE and GPE do not saturate in this limit.  
{The frequencies 
of collective breathing oscillatons in a cigar-shaped trap are 
considered in Sec. \ref{VI}.}
Finally, in Sec. \ref{VII}.
we present the concluding remarks.   

\section{Superfluid hydrodynamics and nonlinear Schr\"odinger equation}
\label{II}

The zero-temperature collective properties of a dilute bosonic 
superfluid under the external potential $U({\bf r})$ 
can be described by the quantum hydrodynamic equations 
\cite{stringa,volovik}: 
\beqa
{\partial n\over \partial t} &\!+\!& \nabla \cdot (n {\bf v}) = 0 \, ,
\label{sf-1}
\\
m {\partial {\bf v}\over \partial t} &\!+\!& \nabla
\left[ -{\hbar^2\over 2m}{\nabla^2 \sqrt{n}\over \sqrt{n} } +
{m\over 2}v^2 + U + \mu(n,a) \right] = 0 \, , 
\label{sf-2}
\eeqa
where $n({\bf r},t)$ is the   local density and 
${\bf v}({\bf r},t)$ is the local velocity of the bosonic 
system. In these zero-temperature hydrodynamical equations
statistics enters the equation of state through the bulk 
chemical potential $\mu(n,a)$ and
the quantum-pressure term $-[\hbar^2/(2m \sqrt{n})]\nabla^2\sqrt{n}$,
which is absent in the classical hydrodynamic equations \cite{landau2}. 
This hydrodynamic regime is achieved in the limit of very   large number 
of atoms $N$.
At zero temperature, the bulk chemical potential $\mu(n,a)$ of the system 
is a function of the density $n$ and of the 
inter-atomic scattering length $a$. For a superfluid 
the velocity field is irrotational, i.e. 
$
\nabla \wedge {\bf v}= 0 \; ,  
$
and the circulation is quantized, i.e. 
\beq 
\oint {\bf v} \cdot d{\bf r} = 2 \pi L {\hbar \over m} \; , 
\eeq
where $L$ is an integer (angular momentum) quantum number \cite{annett}. 

We can introduce the complex order parameter \cite{landau,rmp}
$
\Psi({\bf r},t)=n({\bf r},t)^{1/2} e^{iS({\bf r},t)} 
$ 
such that the phase $S({\bf r},t)$ 
of the order parameter fixes  the superfluid velocity field
\beq
n({\bf r},t) {\bf v} ({\bf r},t)=-i\frac{\hbar}{2m}\left(\Psi^*\nabla 
\Psi-\Psi \nabla\Psi^*\right)
\eeq
so that
${\bf v}({\bf r},t) = ({\hbar/ m})
\nabla S({\bf r},t).$ 
In this way we can map Eqs. (\ref{sf-1}) and (\ref{sf-2}) into
the following time-dependent 
highly nonlinear Schr\"odinger equation  
\beq 
i \hbar {\partial \over \partial t} \Psi 
= \Big[ -{\hbar^2\over 2m}\nabla^2 + U({\bf r}) 
+ \mu(n,a) \Big] \Psi \; , 
\label{mgpe} 
\eeq
In general, one can use the 
hydrodynamic equations (\ref{sf-1}) and (\ref{sf-2}), 
or equivalently Eq. (\ref{mgpe}), 
to study the global properties of the superfluid, 
like the stationary density profile, the free expansion 
and the collective oscillations. 

For a Bose gas the following two leading terms of the low-density 
expansion 
of the bulk chemical potential can be obtained \cite{polls3} from the 
expression for   energy per particle as obtained by 
Lee, Yang and Huang \cite{huang} 
\beq
\mu(n,a) = {4\pi\hbar^2 \over m} a n 
\left( 1 + {32\over 3\pi^{1/2}} (n^{1/3} a)^{3/2} 
+ \ ... \ \right)  \; , 
\label{chemical-low} 
\eeq
where $n^{1/3} a$ is the dimensionless gas parameter \cite{landau}. 
Note that in this expansion the scattering length $a$ must 
be positive ($a>0$) corresponding to a repulsive interaction. Higher 
order correction terms to the bulk chemical potential have also been 
considered in the literature \cite{not}.
The lowest order term of 
expansion (\ref{chemical-low}) was derived by Lenz \cite{lenz}. 
Considering  only this term, Eq. (\ref{mgpe}) becomes 
the familiar GPE \cite{gross,rmp}
\beq 
i \hbar {\partial \over \partial t} \Psi 
= \Big[ -{\hbar^2\over 2m}\nabla^2 + U({\bf r}) 
+ \frac{4\pi\hbar^2}{m}a |\Psi|^2\Big] \Psi \; . 
\label{gpe} 
\eeq
By taking into account also the second term of the expansion 
(\ref{chemical-low}), Eq. (\ref{mgpe}) becomes 
the so-called modified Gross-Pitaevskii equation (MGPE) 
introduced by Fabrocini and Polls \cite{polls1}:
\begin{eqnarray} 
i \hbar {\partial \over \partial t} \Psi 
&=& \Big[ -{\hbar^2\over 2m}\nabla^2 + U({\bf r}) 
+ \frac{4\pi\hbar^2}{m}a |\Psi|^2  \nonumber \\ 
&\times&\biggr(1+ \frac{32}{3\sqrt 
\pi} a^{3/2}|\Psi|
\biggr)
 \Big] \Psi \; . 
\label{mgpe2} 
\end{eqnarray}

In the unitarity limit, where $a\to +\infty$, for dimensional 
reasons \cite{cowell,heis} the bulk chemical potential must be of the 
form  \cite{cowell}
\beq 
\mu(n,a) = \xi {\hbar^2 \over m} n^{2/3} 
\label{chemical-unitarity}
\eeq
where $\xi$ is a universal coefficient. Thus, 
in the unitarity limit the bulk chemical potential is proportional to 
that of a 
non-interacting Fermi gas \cite{skafm}. Recent numerical calculations 
based on Jastrow variational wave functions 
give $\xi = 22.22$ \cite{cowell} and we shall consider this value of 
$\xi$ in 
our calculation. 

In the full crossover from the small-gas-parameter regime 
to the large-gas-parameter regime, we suggest the following 
expression as 
 the bulk chemical potential 
of the Bose superfluid  
\beq 
\mu(n,a) = {\hbar^2\over m} n^{2/3} \ f(n^{1/3}a) 
\label{chemical-general} 
\eeq
where $f(x)$ is an unknown dimensionless universal function
of the gas parameter $x=n^{1/3}a$. 
A general Pad\`e approximant for the function $f(x)$ consistent with the 
expansion of Lee, Huang, and Yang (\ref{chemical-low}) \cite{huang} 
and Eq. (\ref{chemical-general})
should have the following form
\beq 
f(x) = 4\pi { x + \alpha x^{5/2} \over 1 + \gamma x^{3/2}+\beta x^{5/2} 
} \; , 
\label{pade} 
\eeq
where $\alpha$, $\beta$, and $\gamma$ are yet undetermined parameters. 
Without further information about $\mu(n,a)$ we cannot determine all 
these parameters consistently. In this paper  we consider the minimal 
form of this 
function consistent with Eqs. (\ref{chemical-low}) and 
(\ref{chemical-general}) obtained by setting $\gamma =0$ and 
$\alpha = 32/(3 \sqrt{\pi})$ and 
$\beta = 4\pi\alpha/\xi$, with $\xi =22.22$. 
The function $f(x)$ of Eq. (\ref{pade}) 
is such that $f(x)=4\pi [x + 32x^{5/2}/(3 \sqrt{\pi})]$ 
for $x\ll 1$ and the present model reduces to the MGPE 
(\ref{mgpe2})
\cite{polls1}. 
In the  opposite extreme $x\gg 1$, $f(x)=\xi$, and the present 
model reduces to the unitarity limit (\ref{chemical-unitarity}). 
We call Eq. (\ref{mgpe}) equipped with 
Eqs. (\ref{chemical-general}) 
and (\ref{pade}), e.g., 
\begin{eqnarray} 
i \hbar {\partial \over \partial t} \Psi 
&=& \Big[ -{\hbar^2\over 2m}\nabla^2 + U({\bf r}) 
+ \frac{4\pi\hbar^2}{m}a |\Psi|^2  \nonumber \\ 
&\times&\biggr(  \frac{1+ \alpha a^{3/2}|\Psi|}{1+\beta a^{5/2}
|\Psi|^{5/3}}
\biggr)
 \Big] \Psi \; . 
\label{use} 
\end{eqnarray}
by the name 
USE, i.e. unitarity Schr\"odinger equation. As, by construction,  the 
USE 
has the proper weak \cite{polls3}  and strong \cite{cowell} coupling 
limits,
it is appropriate for the study of weak-to-strong coupling crossover.  
This is the 
model equation we use in the following sections to study quantized 
vortices in a Bose superfluid from weak to strong coupling. In this Sec. 
$\Psi$ is normalized by
$\int |\Psi|^2 d{\bf r}=N$.

It is important to stress that the present approach, 
based on the quantum hydrodynamics, 
could be applied to both superfluid bosons and fermions 
\cite{stringa,lipparini}. 
This 
{hydrodynamic} approach is a time-dependent local 
density approximation
with gradient corrections. In the last few years we have 
succesfully applied it to investigate the collective properties 
of different dilute systems, like Bose-Fermi mixtures 
\cite{sala-hydro}, the superfluid Fermi 
gas in the BCS-BEC crossover \cite{manini}, 
and the 1D Lieb-Liniger liquid \cite{sala-lieb}. 
{Here 
we have obtained the USE valid from weak-coupling to unitarity based on 
the general quantum hydrodynamical scheme.}

\section{Quantized vortices} 
\label{III}

The structure of vortices in superfluid $^4$He at zero-temperature 
was investigated many years ago by Chester, Metz and Reatto 
\cite{reatto} by using a many-body variational 
wave function, and more recently by Dalfovo \cite{dalfovo} 
by using the Orsay-Trento density-functional \cite{trento}. 
The two approaches, which give very similar results, 
are based on the assumption that the superfluid velocity 
of the quantized vortex rotating around the cylindric $z$ axis 
is given by 
\beq 
{\bf v} = {\hbar \over m} {L\over \rho} {\bf u}_{\phi}\; , 
\label{vel-vortex}
\eeq
where $L$ is the quantum number of circulation, 
$\rho$ is the cylindric radial coordinate and 
${\bf u}_{\phi}$ is the unit azimuthal vector, 
with $\phi$ the azimuthal angle. 
In 1997 Sadd, Chester and Reatto \cite{reatto2} suggested 
that the velocity of superfluid $^4$He 
is not truly singular at the vortex line and the vorticity 
is distributed over a finite region. Nevertheless, 
the deviations from Eq. (\ref{vel-vortex}) in $^4$He
seem to be very small \cite{reatto2}. 

We recall that there are remarkable differences 
between superfluid $^4$He and the dilute superfluid we are 
considering here. In $^4$He the effective radius $R_0$ 
of the inter-atomic potential is of the order of the average 
distance $n^{-1/3}$ between atoms, i.e. $R_0 n^{1/3} \simeq 1$. 
Instead, in dilute ultra-cold gases the effective radius $R_0$ 
is always much smaller than the average distance 
$n^{-1/3}$ between atoms, i.e $R_0 n^{1/3} \ll 1$. 
For a dilute gas the coupling regime depends on the 
scattering length $a$, namely on the gas parameter $a n^{1/3}$: 
in the weak-coupling regime 
$a n^{1/3} \ll 1$, while in the strong-coupling 
regime $a n^{1/3} \gg 1$. Another remarkable 
difference between liquid helium and quantum gases of alkali-metal 
atoms is the width of the vortex core. The core of a quantized vortex is 
only few angstroms in superfluid $^4$He while it is of the order of 
sub-micron in a dilute atomic gas \cite{tsubota}. 

In our USE we get Eq. (\ref{vel-vortex}) by setting 
$
\Psi({\bf r},t)= \psi(\rho,z) \exp [i
(L\phi-{\mu_0 \over \hbar}t)] \; , 
$ 
where $\mu_0$ is the chemical potential of the inhomogeneous 
superfluid, fixed by the normalization.  
In this way, Eq. (\ref{mgpe}) with Eq. (\ref{chemical-general}) 
becomes 
\beq 
\Big[ -{\hbar^2\over 2m}
\left( \nabla_{\rho}^2 - {L^2\over \rho^2} 
+ {\partial^2\over \partial z^2} \right) + U(\rho,z) 
+ \mu(n,a) \Big] \psi = \mu_0 \ \psi \; ,  
\label{mgpe-vortex}
\eeq
where $\nabla_{\rho}^2=
{1\over \rho}{\partial\over \partial \rho}
(\rho {\partial\over \partial \rho})$ 
is the Laplacian operator in the radial direction 
and ${\hbar^2 L^2\over 2 m \rho^2}$ is the centrifugal term 
which determines the size of the vortex core, that 
is of the order of the healing length
$l_h =\hbar L/\sqrt{2m\mu_0}$
{\cite{natalia}.}
 This expression is 
obtained 
by equating the centrifugal term to the chemical 
potential $\mu_0$. 

We introduce scaled variables by using the characteristic 
length $l_c$ of the system. 
In particular, 
we make the following transformations: 
$\bar{\rho}=\rho/l_c$, 
$\bar{z} = z/l_c$, 
$\bar{a} = a/l_c$, 
$\bar{U} = U (m l_c^2)/\hbar^2$,
$\bar{\mu}_0 = \mu_0 (m l_c^2)/\hbar^2$.
$\bar{\psi}= \psi\ l_c^{3/2}$. 
In this way Eq. (\ref{mgpe-vortex}) becomes
\beqa 
&&
\biggr[ -{1\over 2} 
\frac{\partial^2}{\partial \bar{\rho}^2}-\frac{1}{2\bar{\rho}} 
\frac{\partial}{\partial \bar{\rho}} 
- {1\over 2}\frac{\partial^2}{\partial \bar{z}^2} 
+\frac{L^2}{2\bar{\rho}^2} 
+ \bar{U}(\bar{\rho},\bar{z}) 
\nonumber \\
&& 
+ \bar{\psi}^{4/3} 
f(\bar{\psi}^{2/3} \bar{a}) 
\biggr] \bar{\psi}  = \bar{\mu}_0 \ \bar{\psi}  
\label{mgpe-vortex-scaled} \; ,  
\eeqa
where $\bar{\psi}(\bar{\rho},\bar{z})$ is assumed 
to be real. 

\begin{figure}[tbp]
\begin{center}
{\includegraphics[width=.8\linewidth]{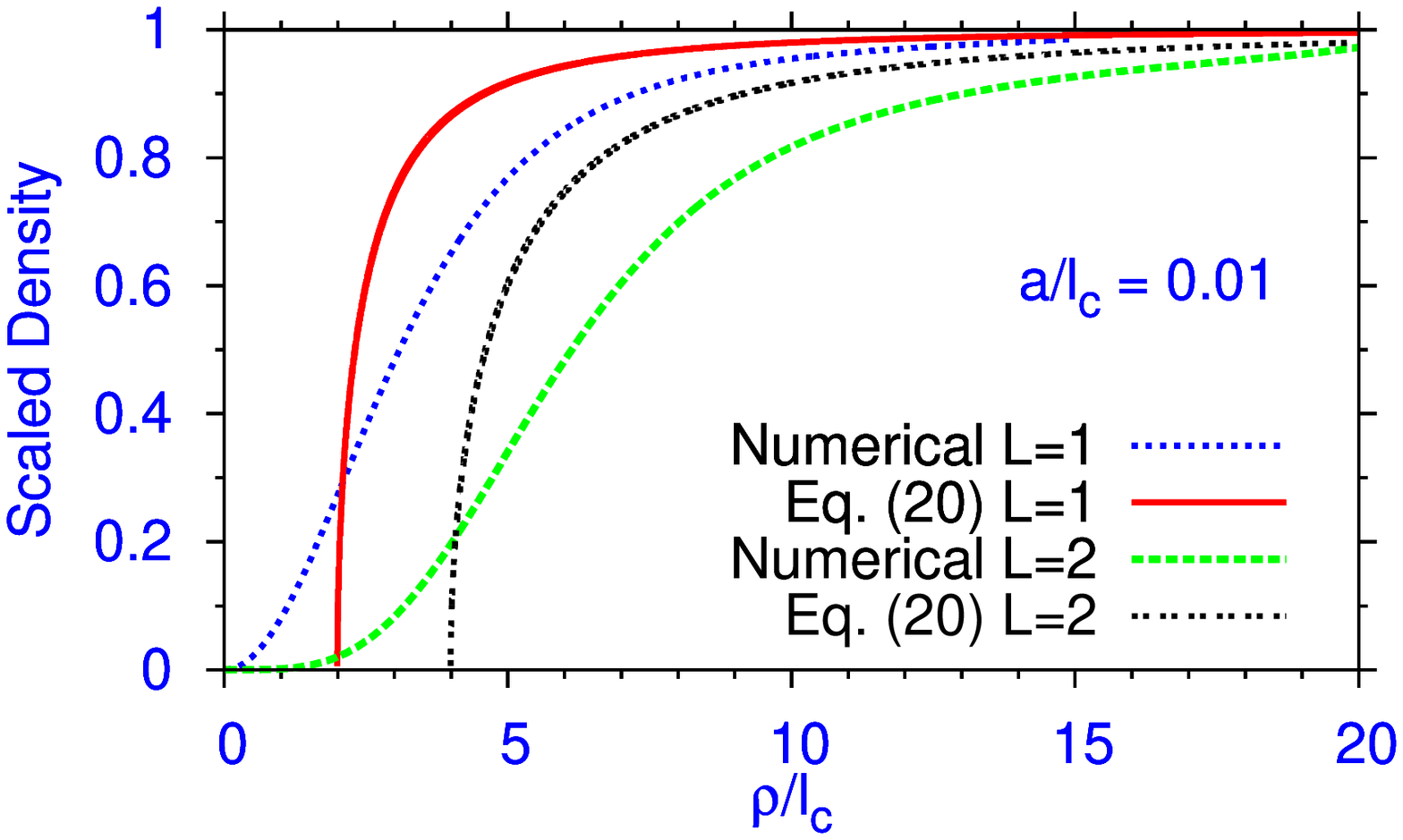}} {%
\includegraphics[width=.8\linewidth]{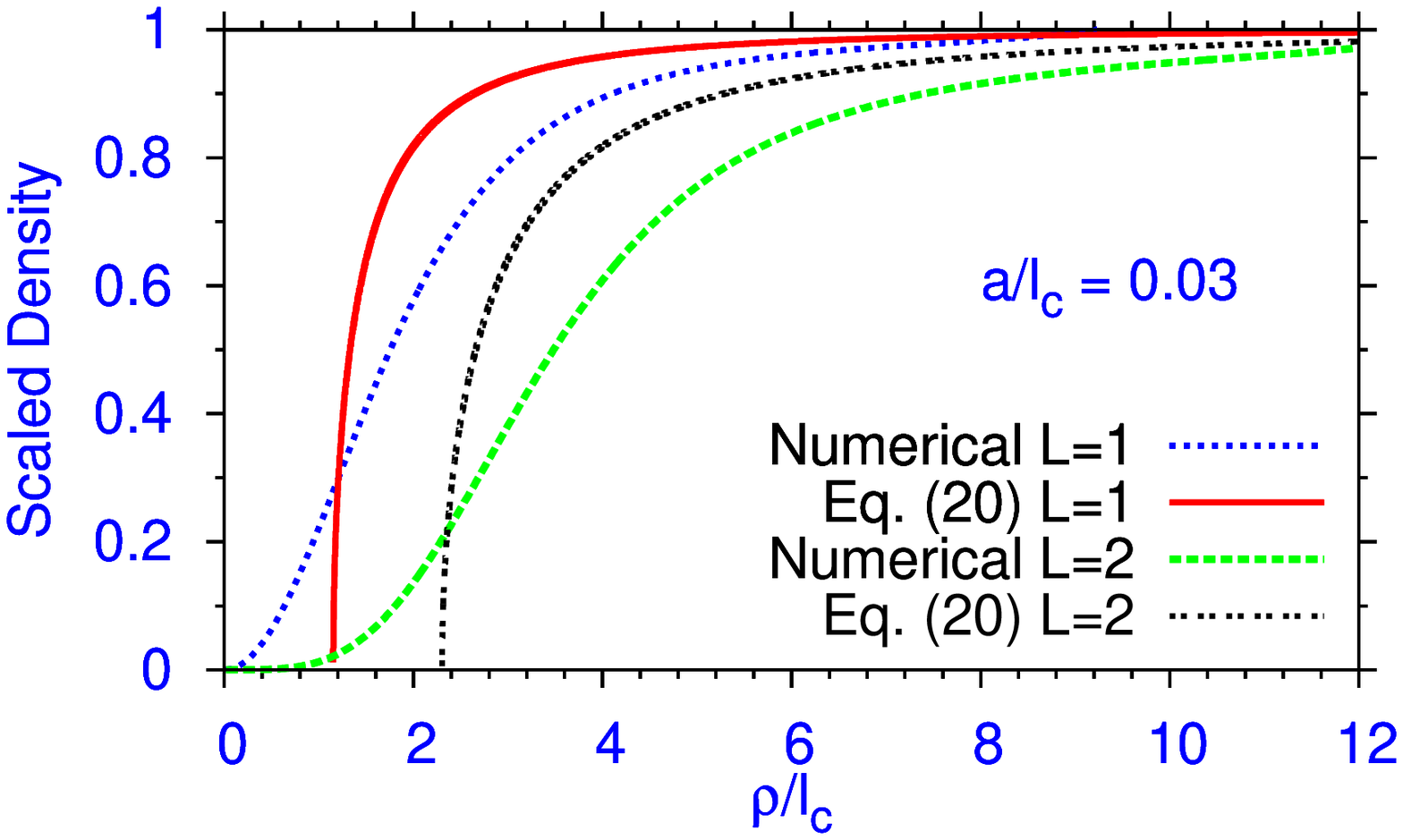}} {%
\includegraphics[width=.8\linewidth]{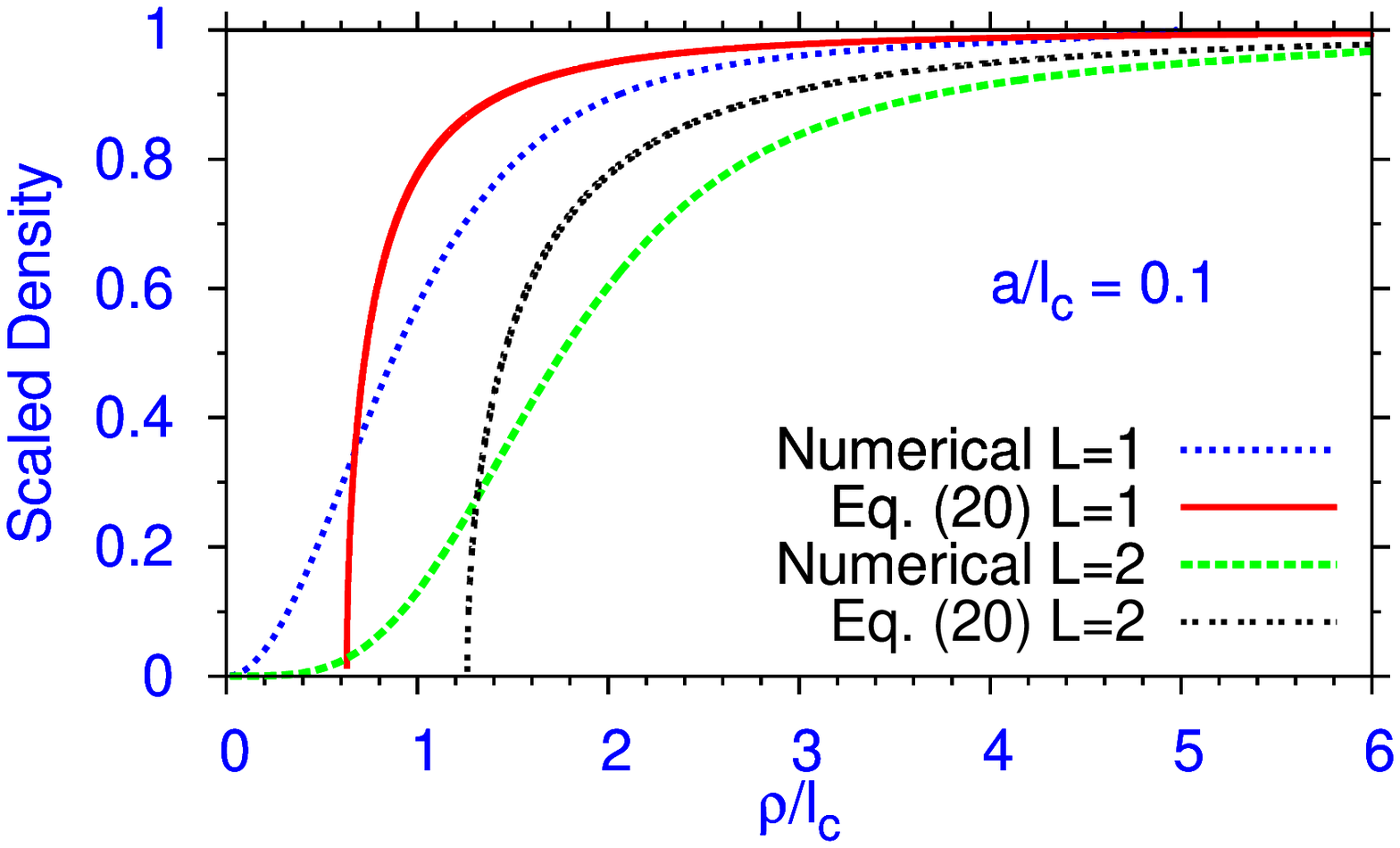}} {%
\includegraphics[width=.8\linewidth]{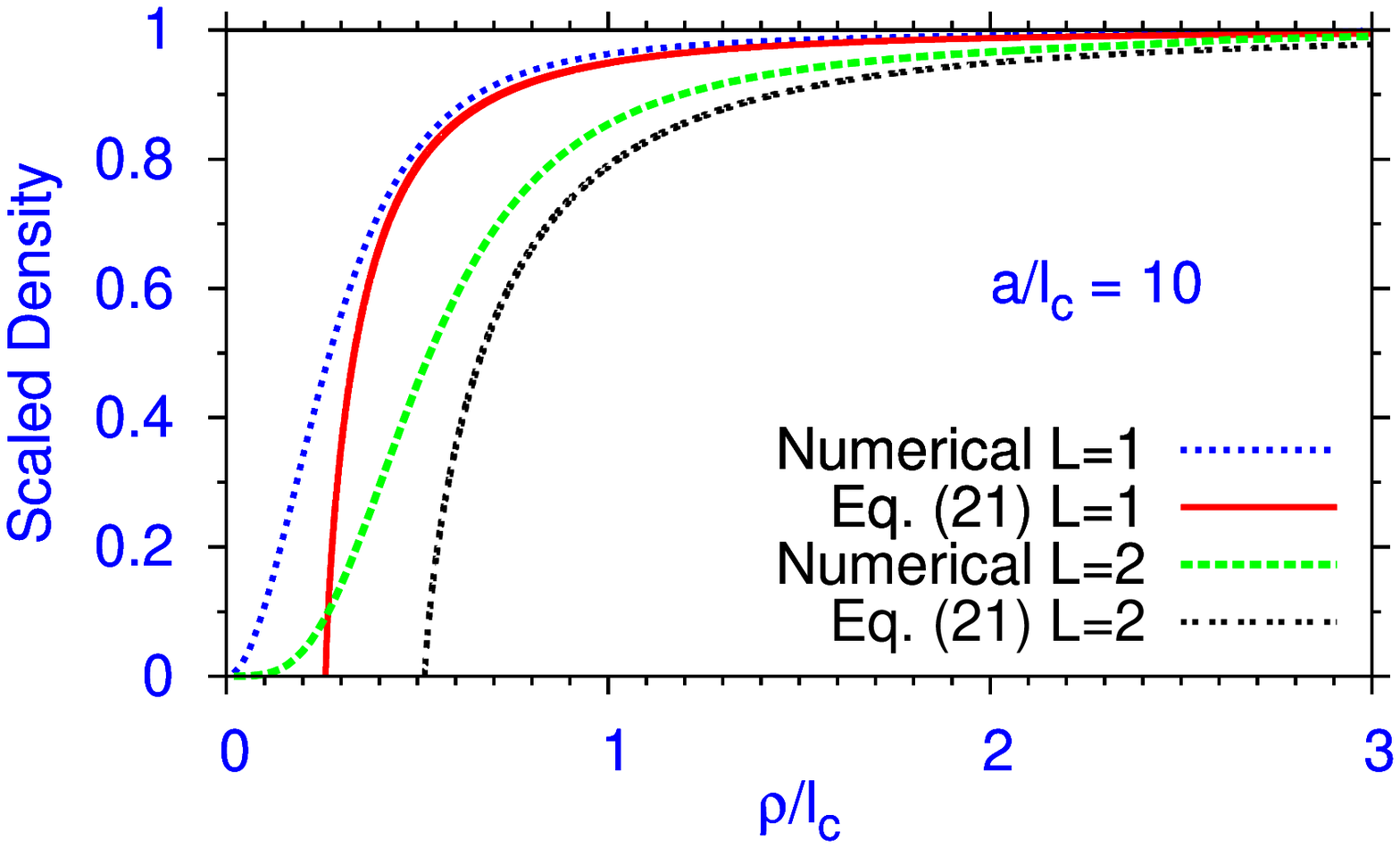}} {%
}
\end{center}
\caption{(Color online) Numerical results
 for 
scaled density of vortices $\bar \psi^2(\bar \rho)$ with $L=1$  and 2 
in a uniform Bose gas obtained by solving the  USE
for  scaled  scattering lengths $a/l_c = $ (a) 
0.01, (b) 0.03, (c) 0.1, and (d) 10
compared with the analytic asymptotic results (\ref{bc2}) or 
(\ref{bc3}).}
\label{fg1}
\end{figure}

\section{Vortex in a uniform Bose gas} 
\label{IV}

Let us first consider the case $U(\rho,z)=0$. 
We suppose that the Bose gas is asymptotically uniform, i.e. 
\beq 
\bar{\psi}(\bar{\rho},\bar{z}) \to 1
\eeq
for $\bar{\rho},\bar{z}\to\infty$. This condition fixes 
the characteristic length $l_c$, that must be 
$l_c= n_{\infty}^{-1/3}$ with $n_{\infty}$ 
the uniform density at infinity. 
In this case the chemical potential is fixed 
by the asymptotic condition, from which one finds 
\beq 
\bar{\mu}_0 = f(\bar{a}) \; . 
\eeq 
In addition, we can choose a wave function which depends only 
on $\bar{\rho}$, and Eq. (\ref{mgpe-vortex-scaled}) becomes 
\beq 
\biggr[ -{1\over 2}
\frac{\partial^2}{\partial \bar{\rho}^2}-\frac{1}{2\bar{\rho}}
\frac{\partial}{\partial \bar{\rho}} 
+\frac{L^2}{2\bar{\rho}^2} 
+ \bar{\psi}^{4/3} 
f(\bar{\psi}^{2/3} \bar{a}) - f(\bar{a}) 
\biggr] \bar{\psi}  = 0 \; . 
\label{mgpe-vortex-scaled-simply}
\eeq
This second-order ordinary differential equation 
must be solved numerically. Clearly with $L\neq 0$ 
the vortex function $\bar{\psi}(\bar{\rho})$ 
has a core around $\bar{\rho}=0$. 
The asymptotic behavior of $\bar{\psi}(\bar{\rho})$ 
for $\bar{\rho}\to \infty$ is obtained by neglecting 
the spatial derivatives in Eq. (\ref{mgpe-vortex-scaled-simply}). 
In this way we obtain the algebraic equation 
\beq 
\bar{\psi}(\bar{\rho})^{4/3} 
f(\bar{\psi}(\bar{\rho})^{2/3} \bar{a}) 
= f(\bar{a}) - \frac{L^2}{2\bar{\rho}^2} \; . 
\label{algebric}
\eeq 
In the weak-coupling regime, $\bar{a}\ll 1$ 
where $f(x) = 4\pi x$, Eq. (\ref{algebric}) gives 
\beq \label{bc2}
\lim_ {\bar{\rho}\to \infty}
\bar{\psi}(\bar{\rho}) = 
\left( 1 - \frac{L^2}{8\pi\bar{a}\bar{\rho}^2} \right)^{1/2} 
 \; . 
\eeq
Instead, in the strong-coupling regime, 
$\bar{a}\gg 1$ where $f(x) = \xi$, Eq. (\ref{algebric}) gives
\beq \label{bc3}
\lim_ {\bar{\rho}\to \infty}
\bar{\psi}(\bar{\rho}) = 
\left( 
1 - \frac{L^2}{2\xi \bar{\rho}^2} 
\right)^{3/4} \; . 
\eeq
Although the limit (\ref{bc2}) depends on the scaled scattering length 
$\bar a$, the limit (\ref{bc3}) is independent of $\bar a$. 
The scaled healing length $\bar{l}_h=l_h/l_c$, which 
estimates the size of the vortex core 
{\cite{natalia},}
is given by 
\beq
\bar{l}_h =L/\sqrt{2f(\bar{a})}. \label{heal} 
\eeq

\begin{figure}[tbp]
\begin{center}
{\includegraphics[width=\linewidth]{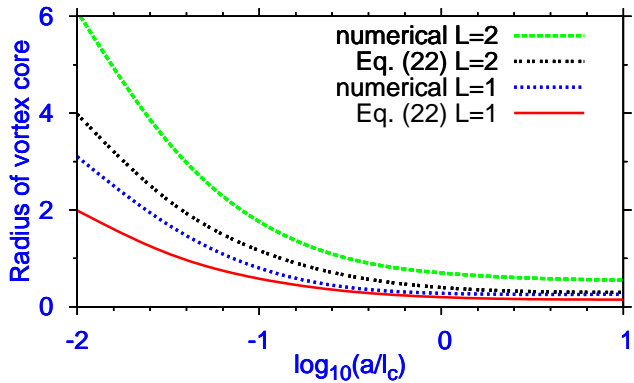}} {%
}
\end{center}
\caption{(Color online). The numerical values of  
vortex core radius for $L=1$ and 2 vs. scaled scattering length 
calculated by the USE
compared 
with 
the analytic results of healing length given by Eq. (\ref{heal}).}
\label{fg2}
\end{figure}

Now we consider 
Eq. (\ref{mgpe-vortex-scaled-simply}) in the limit $\bar \rho \to 0$. 
Because of the regularity of the wave function and the function $f(x)$ 
in this limit the last two terms [the terms 
involving the function $f(x)$] 
in this equation remains finite and can be neglected in comparison to 
the angular momentum term $L^2/(2 \bar \rho^2)$. Consequently, we have 
the following condition at small $\bar \rho$
\begin{equation}\label{bc1}
\lim_ {\bar{\rho}\to 0}
\bar{\psi}(\bar{\rho}) =\bar \rho ^L \times \varphi(\bar \rho),
\end{equation} 
where $\varphi(\bar \rho)$ is a smooth function. 

\begin{table}
\caption{The parameters for $L=1$ used in the numerical solution of Eq. 
(\ref{mgpe-vortex-scaled-simply}) 
as plotted in Fig. \ref{fg1}; in this case $\bar \psi(0) =0$.
For $L=2$,  $\bar 
\psi(0)=\bar\psi'(0)=0$.}
\label{tb1}
\begin{tabular}{|l|l|l|l|}
\hline
$\bar a$ & $\bar\psi'(0)$ & $f(\bar a)$ & $\bar l_h= 1/\sqrt{2f(\bar 
a)}$ 
\\ \hline
 0.01 & 0.2934232 & 0.126415651 &  1.989 \\ \hline
 0.03 & 0.5157240 & 0.388573686 &  1.134  \\\hline
 0.1& 1.0186620 & 1.47985617 & 0.581 \\ \hline
 10 & 3.4622586006 & 22.3160243 & 0.149 \\ \hline
\end{tabular}
\end{table}

We next solve Eq. (\ref{mgpe-vortex-scaled-simply}) for $L=
1$  and 2 using the 
classic fourth-order Runge-Kutta method \cite{RK,num}.  This method  
produces very accurate convergence. We employ a space step of 
$\Delta =0.0001$ and 
integrate up to a scaled distance of $\bar \rho_{\mbox{max}}=20$. The 
integration is started with 
the initial boundary condition (\ref{bc1}) with a trial 
$\bar \psi
(0)$ and 
$\bar \psi 
'(0)$. For $L=1$ this condition is taken as $\bar \psi
(0)=0$  and $\bar \psi'
(0)=$ constant and the integration started at $\bar \rho =0$. 
For $L=2$ both $\bar \psi
(0)=0$ and $\bar \psi'
(0)=0$ and the numerical integration cannot be started at $\bar \rho 
=0$, as then $\bar \psi$ as well as $\bar \psi'$ at subsequent sites 
become zero.  
For L=2, the integration is started at $\bar \rho=\Delta $ with the 
boundary condition $\bar \psi
(\Delta)=0$ and $\bar \psi'
(\Delta)$ equal to a small constant.      
The integration is 
propagated to $\bar \rho=\bar 
\rho_{\mbox{max}}$ where the asymptotic conditions (\ref{bc2}) or 
(\ref{bc3}) remain valid. If after integration with a trial
guess,  the boundary conditions (\ref{bc2}) or
(\ref{bc3}) cannot be satisfied, the method is implemented with a new 
trial guess. The process is continued until a  solution   satisfying the 
proper boundary conditions at small and large $\bar \rho$ is obtained.  
We obtain the solution for 
different values of $\bar a$. 

The numerical results for the scaled density [$\equiv \bar \psi^2(\bar 
\rho)$] for different scaled scattering length $\bar a \equiv a/l_c$ is 
plotted in Fig. \ref{fg1} and compared with the asymptotic forms 
(\ref{bc2}) or (\ref{bc3}) for $L=1$ and 2. The corresponding parameters  
$\bar \psi 
'(0)$ and $f(\bar a)$ used in the numerical solution are given in Table 
\ref{tb1}.  The quantity $f(\bar a)$ shown in Table \ref{tb1} is 
interesting   as it can be compared with the chemical potential of Eq. 
(\ref{mgpe-vortex-scaled-simply}). Hence the numerical values of $f(\bar 
a)$ for different $\bar a$ should give an idea of the variation of 
chemical potential with coupling.  The quantity $f(\bar a)$ 
is also related to the scaled healing length $\bar l_h$ given by Eq. 
(\ref{heal}) as shown in Table \ref{tb1}. 

It is interesting to calculate numerically the vortex core radius defined 
as the value of $\bar \rho$ for which $\bar \psi^2 (\bar \rho)=0.5$.  
In Fig. \ref{fg2} we plot the numerical values of vortex core radius  
versus scaled scattering length and compare with the corresponding 
analytical results of healing length given by Eq. (\ref{heal}). 
The figure shows that the vortex-core radius has the same trend 
of the analytical healing length 
{\cite{natalia}.} 
From Figs. \ref{fg1} and \ref{fg2} we find that the
vortex-core radius decreases 
with increasing coupling but increases with angular momentum $L$. 
However, for very large coupling it saturates. 
This  saturation of the vortex core is properly given by   the   USE. 
 
\section{Vortex in a Bose gas in a harmonic trap} 
\label{V}

\begin{figure}[tbp]
\begin{center}
{\includegraphics[width=.9\linewidth]{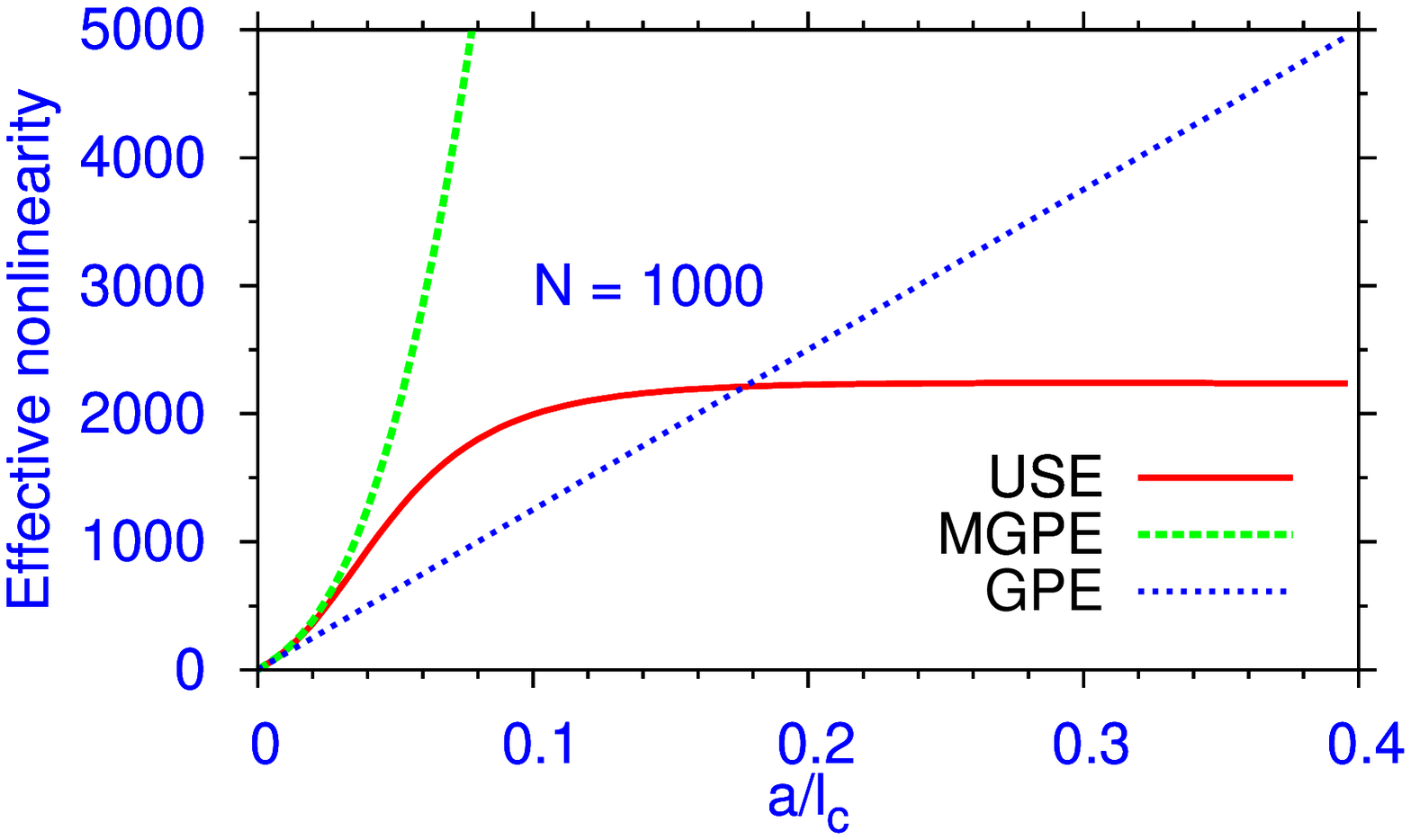}} 
{\includegraphics[width=.9\linewidth]{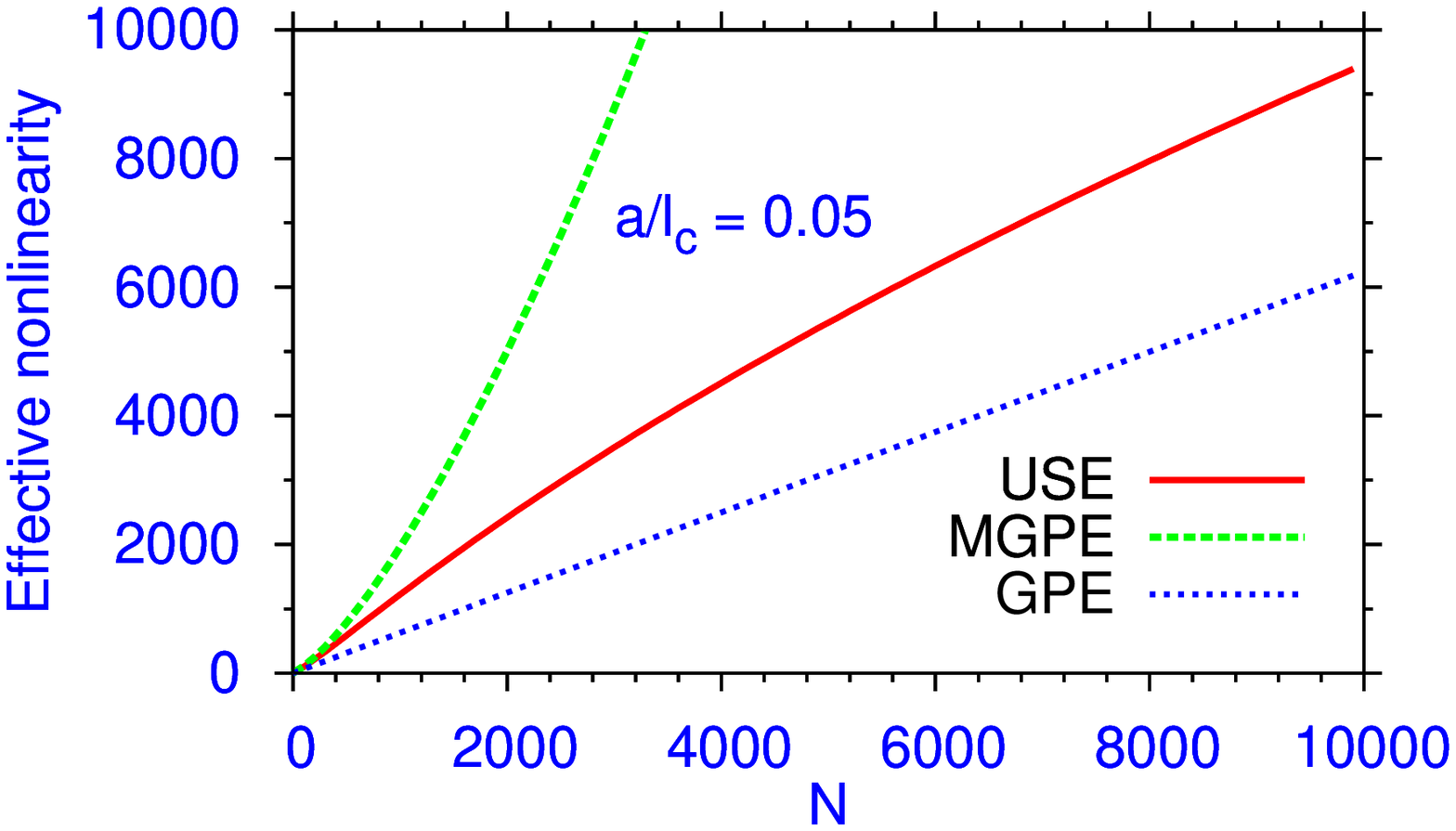}} 
\end{center}
\caption{(Color online). The numerical values of effective 
nonlinearity $N^{2/3} f(N^{1/3}  \bar a)$ of the USE,  
 $4\pi N \bar a$ of the GPE, and  $4\pi N \bar a(1+\alpha \bar 
a^{3/2}\sqrt N)$ of the MGPE  
for $\bar \psi=1$ for (a) $N=1000$ vs. 
$a/l_c$ and (b) $a/l_c=0.05$ vs. $N$.}
\label{fg3}
\end{figure}

Now we consider the bosonic fluid in an axially-symmetric  harmonic 
trapping potential, i.e.
\beq \label{cpot} 
U={1\over 2} m \omega_{\rho}^2 (\rho^2 + \lambda^2 z^2)
\eeq
where $\lambda=\omega_z/\omega_{\rho}$ is the anisotropy 
parameter, with $\omega_{\rho}$ the transverse frequency and 
$\omega_z$ the axial frequency. We choose as characteristic 
length of the system the transverse harmonic length, i.e.  
$l_c=\sqrt{\hbar/(m\omega_{\rho})}$. 
By using the scaled variables the confining potential reads 
$
\bar{U} = {1\over 2} (\bar{\rho}^2 + \lambda^2 \bar{z}^2) \; .  
$
With this confining potential the mean-field equation can be written 
explicitly as
\beqa 
&&
\biggr[ -{1\over 2} 
\frac{\partial^2}{\partial \bar{\rho}^2}-\frac{1}{2\bar{\rho}} 
\frac{\partial}{\partial \bar{\rho}} 
- {1\over 2}\frac{\partial^2}{\partial \bar{z}^2} 
+\frac{L^2}{2\bar{\rho}^2} 
+ {1\over 2} (\bar{\rho}^2 + \lambda^2 \bar{z}^2) 
\nonumber \\
&& 
+ 4\pi N \bar a \bar{\psi}^{2} \frac{1
+\alpha\bar a ^{3/2} 
N^{1/2}\bar \psi}{1+\beta \bar a ^{5/2}
N^{5/6}\bar \psi^{5/3}
} 
\biggr] \bar{\psi}  = \bar{\mu}_0 \ \bar{\psi} 
\label{eq1} \; ,  
\eeqa
Now the normalization condition is  given by 
\beqa
2\pi \int_0^\infty \bar{\rho} \ d\bar{\rho}
\int_{-\infty}^{\infty} d\bar{z} \
\bar{\psi}^2(\bar{\rho},\bar{z})\equiv 
2\pi \int_0^\infty \bar{\rho} \ d\bar{\rho} n_c(\bar \rho) = 1 \; .
\label{eq2}
\eeqa
Here we defined $n_c(\bar \rho)$ as  the column density (as in 
\cite{polls3}) appropriate for the study of radial distribution of 
matter in the vortex. 
Equations (\ref{eq1}) and (\ref{eq2}) constitute the USE for an axially 
trapped  Bose gas valid from weak-coupling ($\bar a 
\to 0$ ) to the unitarity limit 
($\bar a \to \infty$) and (in spite of a slightly 
complicated numerical structure) it is no more 
difficult to solve than the usual GP equation valid 
in the weak-coupling limit. 
In the extreme weak-coupling limit, the non-linear term of Eq. (\ref{eq1}) 
becomes   the usual GP term $4\pi N \bar a \bar \psi^2$. In the 
unitarity limit this term becomes independent of scattering length 
and equals $\xi N ^{2/3}\bar \psi^{4/3}$, $\xi = 
22.22$.  We recall that if we use the GPE (\ref{gpe}), as well as 
the MGPE (\ref{mgpe2}), in the unitarity limit, it will lead 
to an inappropriate dependence of the nonlinearity on scattering 
length, as well as on the number of atoms $N$. 

\begin{figure}[tbp]
\begin{center}
{\includegraphics[width=\linewidth]{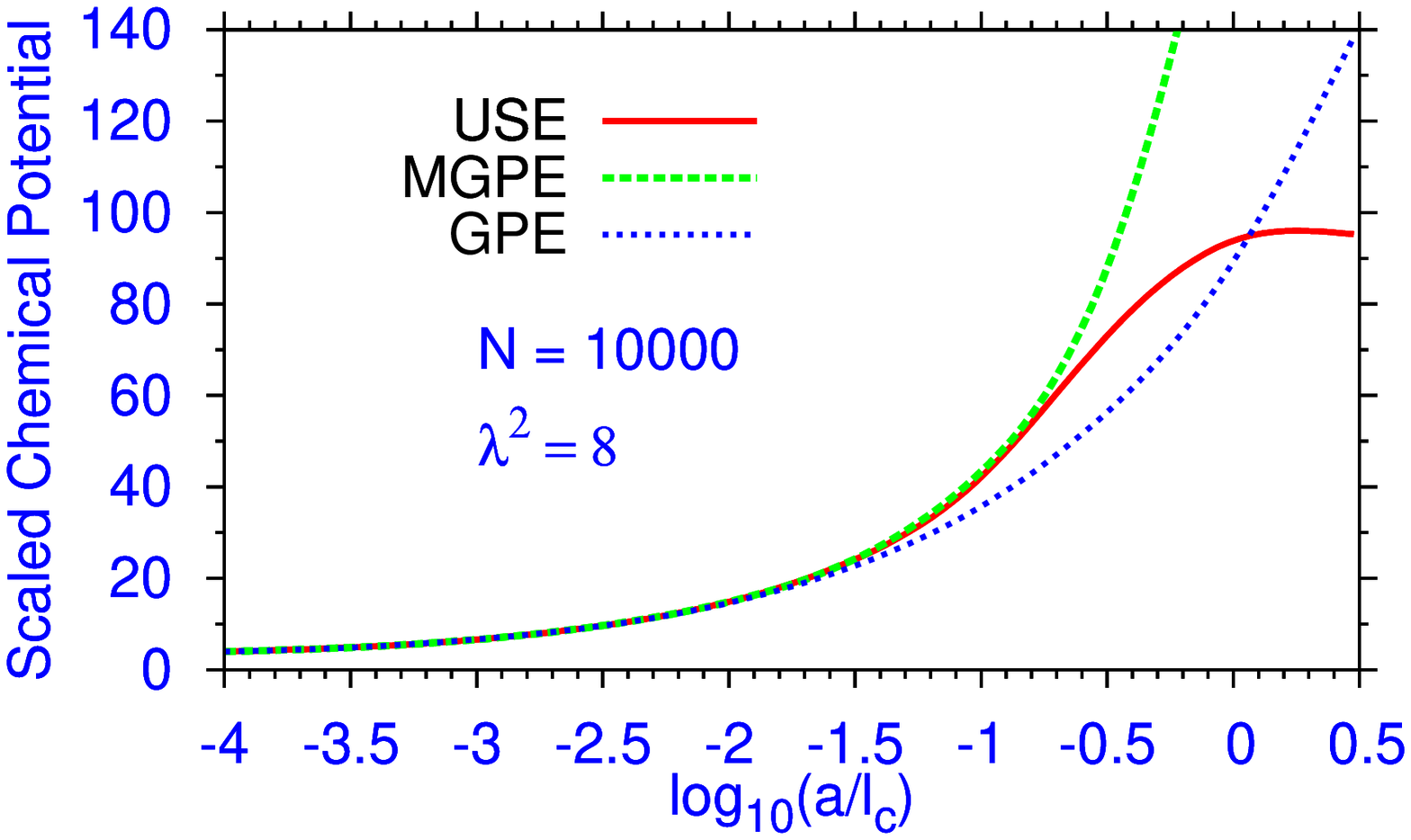}} 
{\includegraphics[width=\linewidth]{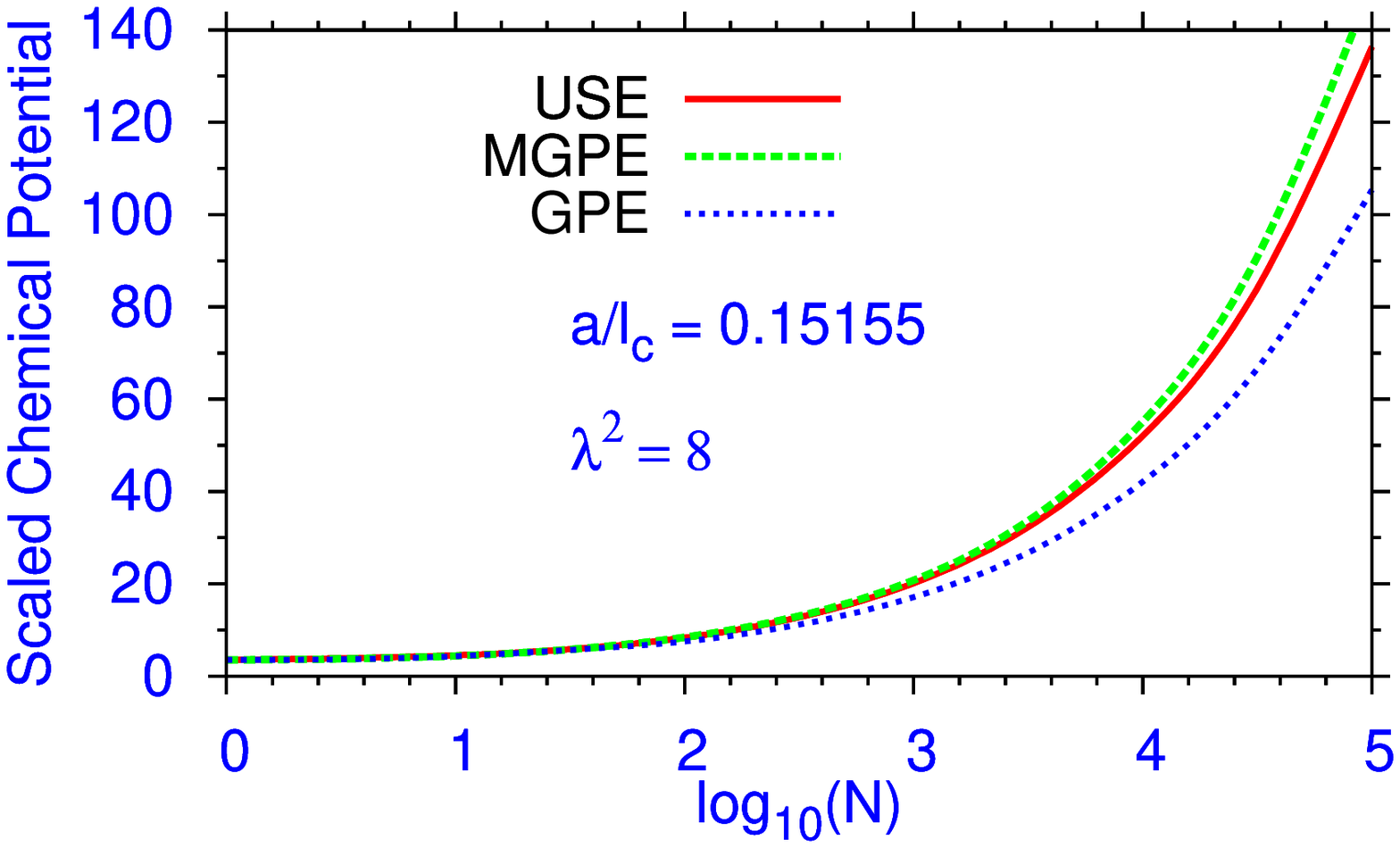}} 
\end{center}
\caption{(Color online). The numerically calculated 
scaled chemical potential $\bar \mu_0$ (in 
units of 
oscillator energy $\hbar \omega_\rho$) from GPE, MGPE and 
USE vs. (a) scaled scattering 
length $a/l_c$ for $N=10000, \lambda^2=8$, and vs. (b)   
$N$ for $a/l_c=0.15155, \lambda^2=8$. 
}
\label{fg4}
\end{figure}

It is appropriate to study the behavior of the nonlinearity of 
the USE (\ref{eq1}), that is  
$N^{2/3}\bar \psi^{4/3}f(N^{1/3} \bar \psi^{2/3}\bar a)$. 
To study the dependence of this nonlinearity on 
$N$ and $\bar a$ we set 
$\bar \psi=1$ in this expression. The resultant nonlinearities are 
plotted in Fig. \ref{fg3} (a) and (b) as a function of $\bar a $ and $N$ 
for constant $N=1000$ and $a/l_c=0.05$, respectively.   
The interesting feature of the nonlinearity of this equation is 
exhibited in Fig. \ref{fg3} (a), where we find that, for a fixed $N$, 
the nonlinearity of 
this equation saturates at large scattering lengths $\bar a$. In the GP 
 model the nonlinearity increases linearly 
for all $\bar a$, whereas for the MGP model it increases indefinitely, 
but with a more complicated dependence on $\bar a$. From Fig. \ref{fg3} 
(b) we find that for a 
fixed scattering length  the nonlinearity of  Eq. (\ref{eq1}) as well 
as the GP equation increases with $N$. For the GP equation it increases 
linearly with $N$, whereas for Eq. (\ref{eq1}) it increases as $N^{2/3}$
for large $N$. 

To further study the USE (\ref{eq1}), 
we solve it numerically. For numerical convenience we transform 
this equation to its time-dependent form by replacing the term $\bar 
\mu_0\bar \psi$ by its time-dependent counterpart $i {\partial 
\bar \psi}/{\partial t}$  appropriate for non-stationary states. However, 
in this paper we limit ourselves to a study of stationary states. 
Nevertheless, the introduction of time-dependence allows for a solution 
of Eq. (\ref{eq1}) by imaginary time propagation method after 
discretizing it with the semi-implicit Crank-Nicholson rule 
\cite{num,sala-numerics,cn}. In the process of 
discretization we use a time step of 0.0005 and a space step of 0.03. 
We limit our numerical 
study to the $L=1$ vortex, as $L>1$ vortices are unstable and decay to 
several $L=1$ vortices conserving the angular momentum.
In the numerical 
simulation of a vortex with $L=1$ 
we employ a pizza-shaped   condensate with trap anisotropy $\lambda^2 
=8$ 
as in the theoretical study of 
Nilsen {\it et al.} \cite{polls3} and the experiment at JILA 
\cite{jila}. After the wave function $\bar \psi(\bar \rho)$ of Eq. 
(\ref{eq1}) is obtained by the imaginary time propagation, the chemical 
potential is obtained by multiplying this equation by  $\bar \psi(\bar 
\rho)$ and integrating over all space.

In Fig. \ref{fg4} (a) we plot the numerical results of 
scaled chemical 
potential vs. scaled scattering length $a/l_c$ as  
obtained from the USE (\ref{eq1})  as well as the GPE (\ref{gpe})
for $N=10000$. In Fig. \ref{fg4} (b) we plot the same vs. $N$ for 
$a/l_c=0.15155$. The scaled scattering length $a/l_c=0.15155$ is the 
value studied by Nilsen {\it et al.} and is appropriate for Rb atoms in 
an experimental trap used at JILA for a scattering length $a=35 a$(Rb), 
where $a$(Rb)$=100a_0$ ($a_0$ the Bohr radius) is the experimental 
atomic scattering length of Rb. 
In Table \ref{tb2} we plot some of the results for chemical potential 
$\bar{\mu}_0$ for different $\bar a$ and $N$ values. 
The first row in this Table 
agrees with the calculation of Nilsen {\it et al.} for the GP and the 
MGP models. 

{\begin{table}
\caption{Scaled chemical potential $\bar \mu_0$ of GPE, MGPE, and 
USE for different $\bar a=a/l_c$, $N$, and GP nonlinearity $\bar a N$.}
\label{tb2}
\begin{tabular}{|l|l|l|l|l|l|}
\hline
$\bar a$ & $N$ & $\bar a N$        &  
$\mu_{\mathrm{GPE}}$ &   $\mu_{\mathrm{MGPE}}$ & 
$\mu_{\mathrm{USE}}$ 
\\ \hline
 0.15155 & 500 &75.775  &13.19 & 15.62 & 15.27 \\ 
 0.01 & 10000 & 100 & 14.63  &  14.88 & 14.88 \\ 
 0.01 & 100000 &1000 &35.73 &  36.74  &36.70\\ 
 0.01 & 1000000 &10000 & 89.25    & 93.16 &92.99 \\ 
 0.1 & 10000 & 1000 &35.73   &43.32  & 42.10\\ 
 1 & 10000 & 10000  & 89.25    & 196.13 & 93.83\\ 
 3 & 10000 & 30000  & 138.40    & 460 & 95.26 \\ \hline
\end{tabular}
\end{table}
}

\begin{figure}[tbp]
\begin{center}
{\includegraphics[width=\linewidth]{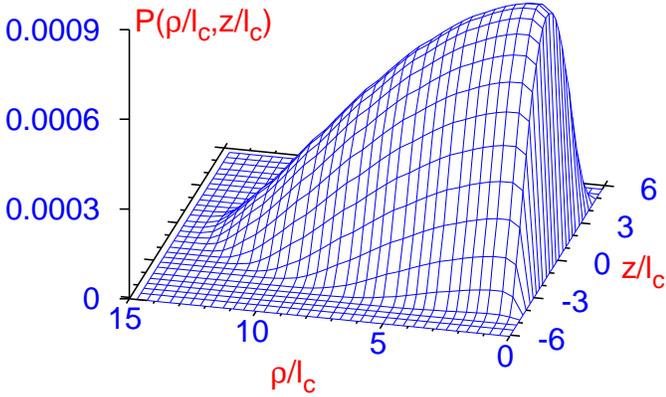}} 
\end{center}
\caption{(Color online)
A typical profile of the probability density 
$P(\rho/l_c,z/l_c)=\bar \psi^2(\rho/l_c,z/l_c)$ 
for a vortex obtained with 
the USE for $N=10000$ and $\bar a=1$.}
\label{fg5}
\end{figure}
 
From Fig. \ref{fg4} (a) and Table \ref{tb2} we find a saturation of the 
present mean-field 
result for large scattering length in the unitarity limit. This is 
consistent with the saturation of the nonlinearity of the USE 
(\ref{eq1}) 
in this limit. After this saturation is obtained, any further increase 
in the scattering length at a fixed $N$ does not change the chemical 
potential $\bar{\mu}_0$. 
Figure \ref{fg4} must be compared with Fig. \ref{fg3}. 
For a fixed scattering length, as $N$ is increased the nonlinearity 
of the USE (\ref{eq1}) keeps on increasing. As a consequence the 
chemical 
potential also increases indefinitely as $N$. In Fig.  \ref{fg4} (b)  
the USE chemical potential is always greater than the GP one.
On the other hand, for a fixed $N$, as $\bar a$ is increased the 
nonlinearity  of the USE saturates above a certain value of $\bar 
a$. This has a consequence in the results: With the 
increase of scattering length, as the system tends to the unitarity limit, 
there are a saturation of the present chemical potential and a 
crossover, beyond which the USE chemical potential becomes 
smaller than the GP one. From Figs. \ref{fg4} we find that  
for moderate to small $\bar a$ values the results for chemical potential 
of the MGPE  (\ref{mgpe2}) \cite{polls1} and USE (\ref{eq1}) 
remain very close to each other. 
From 
Table \ref{tb2} we find that for $N=10^6$, and $\bar a=0.01$ 
(corresponding to a large GP nonlinearity of $N\bar a=10000$) the 
chemical potentials of the MGPE and the USE differ by less than 
$0.2\%$ whereas the chemical potentials of  these two models differ from 
that of the the GP model by about $3\%$.
(We recall that $\bar a=0.01$ is the  typical experimental value of this 
quantity in the experiment at JILA \cite{rmp,jila}.)  
The chemical potentials of MGPE and USE start to differ  
for  large $\bar a$ 
values (larger than the experimental value but attainable by  the 
Feshbach resonance technique \cite{cornish}): the chemical potential of 
the USE model 
exhibits saturation, whereas the chemical potential of the 
MGP model increases very rapidly with increasing $\bar a$ values
which has been made explicit in the data in the last two rows of Table 
\ref{tb2}, where the results of the MGPE \cite{polls1}
and USE show the main 
differences. 

\begin{figure}[tbp]
\begin{center}
{\includegraphics[width=\linewidth]{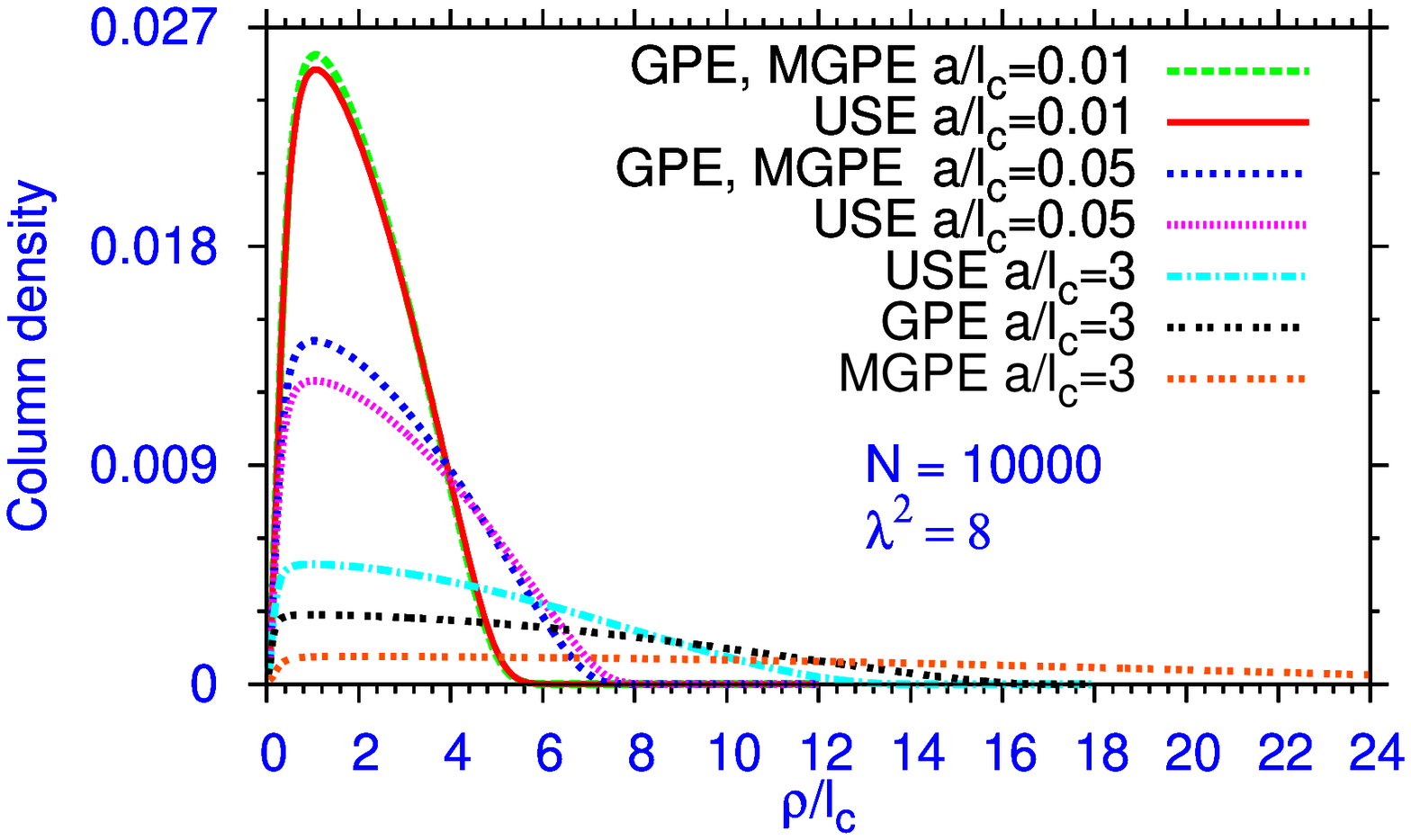}} 
{\includegraphics[width=\linewidth]{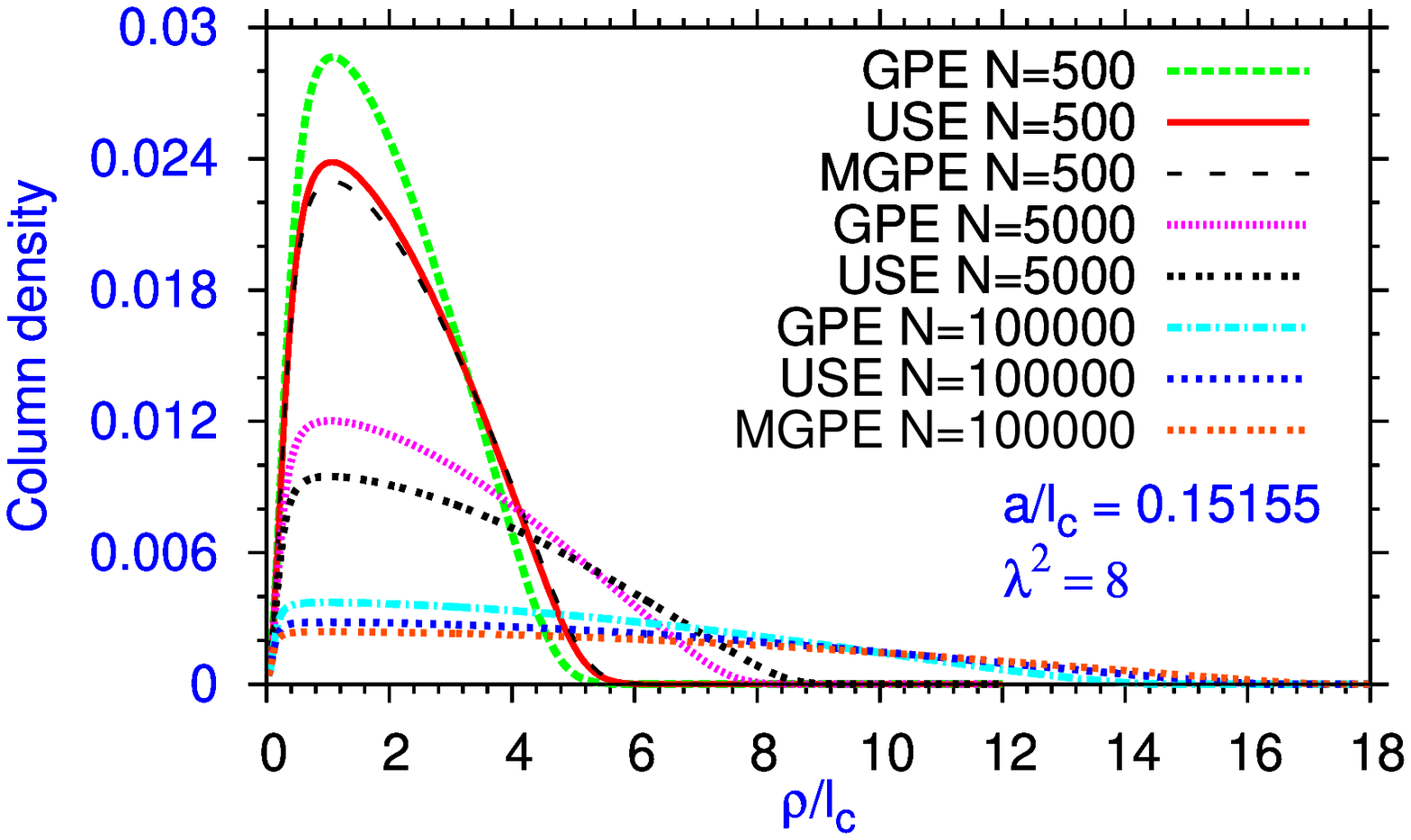}} 
\end{center}
\caption{(Color online). The numerically calculated 
column density $n_c(\bar \rho)= \int_{-\infty}^{\infty}d\bar z \bar 
\psi^2(\bar \rho, \bar z)$ from  
the USE, GPE, and MGPE vs. $\rho/l_c$  for 
 (a)  $N=10000, \lambda ^2 =8$ and  different $a/l_c$, and 
 (b)  $a/l_c=0.15155, \lambda^2=8$ and different $N$. 
}
\label{fg6}
\end{figure}

In Fig. \ref{fg5} we demonstrate a typical profile of the probability 
density of a vortex with $L=1$ as obtained from the USE for 
$N=10000$ and 
$\bar a=1$, corresponding to a large GP nonlinearity of $N\bar a=10000$.
The density is manifestly zero for $\rho=0$ for the vortex state.  
However, both for theoretical and experimental analysis   
it is  appropriate and more convenient to  calculate  the column 
density $n_c(\bar \rho)$ 
as defined in Eq. 
(\ref{eq2})  to 
study the radial distribution of matter in a vortex for the 
USE,  GPE and MGPE. The results for column density 
are plotted in Figs. \ref{fg6} for (a)  $N=10000$ for different $l/l_c$ 
and for (b) $ a/l_c=0.15155$ for different $N$. The density profile  of 
Fig. 
\ref{fg6} (b) for $N=500$ is also reported in \cite{polls3} and the two 
agree with each other for the GP equation.  An interesting 
feature of the plots of Figs. \ref{fg6} is that with an increase of 
nonlinearity (either via an increase of $a/l_c$ for a fixed $N$, or via 
an increase of $N$ for a fixed $a/l_c$), the column density extends up to 
a larger radius. Of the  column densities, that obtained with the 
USE extends to a larger radius than the GP column density in general. This 
trend is reversed in the unitarity limit of large scattering length as 
can be seen from the plot of Fig. \ref{fg6}  (a) for $a/l_c=3$. 
{For 
small values of $a/l_c$ the MGPE results are close to the GPE results. 
For medium values of $a/l_c$ the MGPE results are close to the USE 
results. However, for large values of $a/l_c$ the MGPE wave functions 
extend to very large distances, compared to the other models,  
due to an unphysically  large  
nonlinearity in this model (MGPE). }

\begin{figure}[tbp]
\begin{center}
{\includegraphics[width=\linewidth]{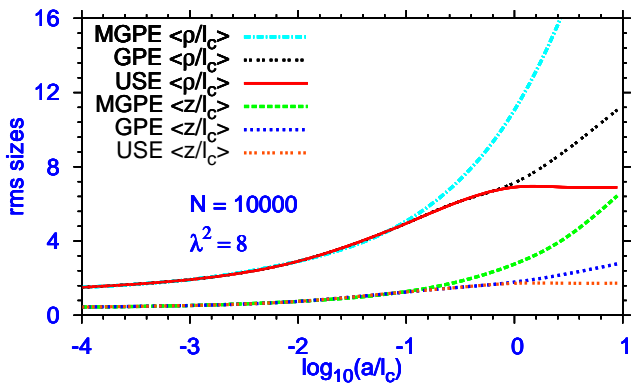}} 
\end{center}
\caption{(Color online). 
The numerically calculated scaled rms sizes $\langle \rho/l_c 
\rangle $ and $\langle z/l_c \rangle$
of the condensate from USE, GPE, and MGPE
vs. 
scaled scattering length $\bar a = a/l_c$ 
}
\label{fg7}
\end{figure}

A manifestation of the saturation of nonlinearity of the USE is explicit 
in the rms (root mean square) radial ($\langle \rho/l_c \rangle $) 
and axial  ($\langle z/l_c \rangle$) sizes of the condensate as the 
scattering length 
$a$ is increased. The results for the rms sizes calculated with the GPE 
keeps on increasing indefinitely as the scattering length is increased. 
The rms sizes calculated with the MGPE increases even more rapidly than 
the results of the GPE. However, the rms sizes calculated with the USE 
saturates after a certain value of $a$, beyond which they remain 
constant. This is illustrated in Fig. \ref{fg7} where we plot the rms 
sizes vs. log$(a/l_c)$ as calculated by the USE, GPE, and MGPE. As we 
are 
considering a pizza-shaped condensate the radial size $\langle  
\rho/l_c \rangle $ is larger than the axial size $z/l_c$. 
The theoretical results for the rms sizes exhibited in Fig. \ref{fg7} 
can be verified experimentally with present technology and this will 
provide a test for the USE proposed in this paper.

\section{Collective oscillations in  harmonic confinement}
{
We consider the effect of confinement due to an external
anisotropic harmonic potential (\ref{cpot})
on frequencies of collective oscillation of the system from 
weak-coupling to unitarity using the USE. 
It has been shown by Cozzini and Stringari 
\cite{cozzini}
that assuming a power-law dependence $\mu = A \; n^{\Gamma}$
for the chemical potential (polytropic equation of state
\cite{combescot}), from Eqs. (\ref{sf-1}) and (\ref{sf-2})
without the quantum pressure term, one finds analytic
expressions for the collective breathing frequencies of the superfluid.
In particular, for the very elongated cigar-shaped traps, 
the collective radial breathing
mode frequency $\Omega_{\rho}$ is given by \cite{cozzini}
\beq
\Omega_{\rho} = \sqrt{2(\Gamma + 1)} \; \omega_{\rho} \; ,
\label{wr}
\eeq
while the collective longitudinal breathing mode $\Omega_{z}$ is
\beq
\Omega_z = \sqrt{3\Gamma +2 \over \Gamma +1} \; \omega_z \; .
\label{wz}
\eeq
}
\begin{figure}[tbp]
\begin{center}
{\includegraphics[width=\linewidth]{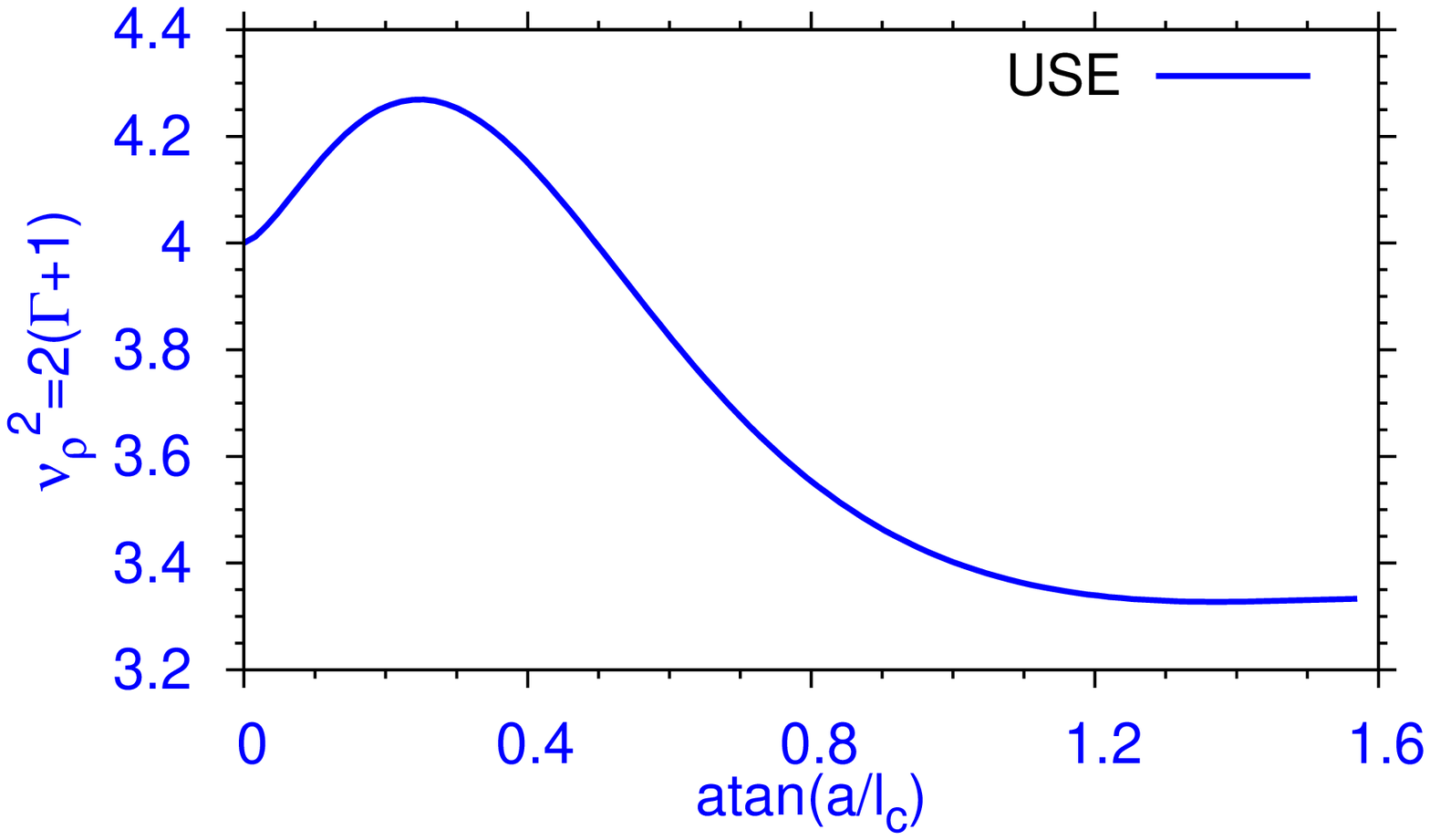}} 
{\includegraphics[width=\linewidth]{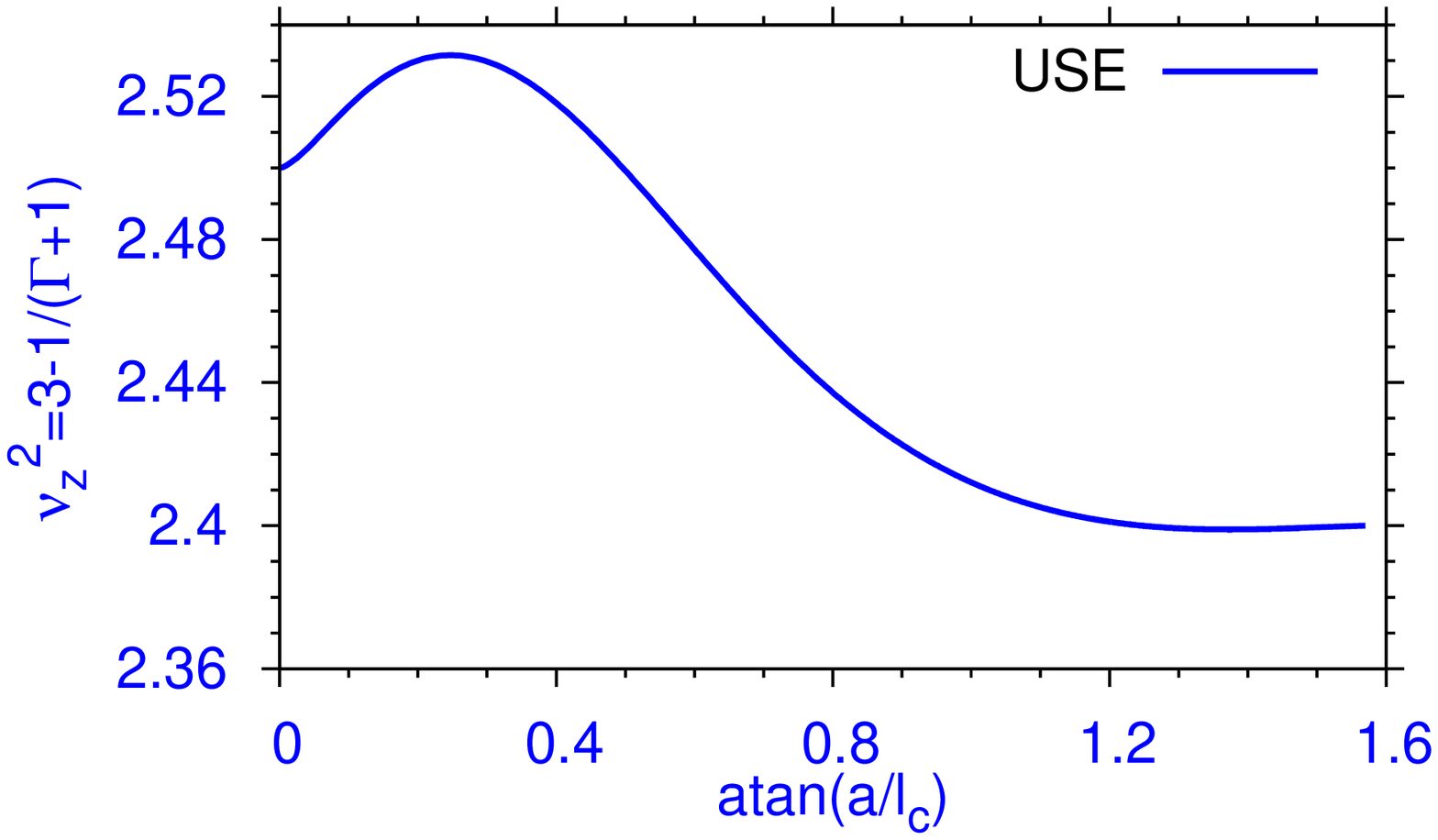}} 
\end{center}
\caption{
{(Color online) Square of the reduced 
frequencies of collective 
radial and axial oscillations (a) $\nu_\rho^2$ and (b) $\nu_z^2$ in a 
cigar-shaped trap vs. atan$(a/l_c).$
}}
\label{fg8}
\end{figure}

{
Here we introduce an effective polytropic index
$\Gamma$ as the logarithmic derivative of the bulk chemical
potential $\mu$, that is
\beq
\Gamma = {n \over \mu} {\partial \mu \over \partial n} =
{2 \over 3} + {1\over 3} x {f'(x)\over f(x)}  \; ,
\label{gamma}
\eeq
where $x=n^{1/3}a$ is the gas parameter.
This formula for the local polytropic equation is useful
to have a simple analytical prediction of the collective frequencies.
In the weak-coupling regime ($x\ll 1$) one finds
$x f'(x)/f(x)= 1$ and $\Gamma=1$, consequently,  the reduced frequency 
square $\nu _\rho^2\equiv  (\Omega_\rho/\omega_\rho)^2= 2(\Gamma 
+1))=4$ and  $\nu _z^2\equiv  (\Omega_z/\omega_z)^2= 3-1/(\Gamma+1)
=5/2$;
while in the unitarity regime
($x \gg 1$) it holds $x f'(x)/f(x)= 0$ and $\Gamma=2/3$ and 
consequently,  $\nu 
_\rho^2=10/3$ and  $\nu
_z^2=12/5$ .
The analytical predictions of Eqs. (\ref{wr}) and (\ref{wz})
with Eq. (\ref{gamma}) are shown in Fig. \ref{fg8}, where we plot 
$\nu_\rho^2$ and $\nu_z^2$ vs. atan$(a/l_c),$ so that the entire region 
$\infty> a \ge 0$ is mapped into the finite interval $\pi/2\ge 
\mbox{atan}(a/l_c)  \ge 0$.
}

\label{VI}

\section{Conclusion} 
\label{VII}

In this paper we have proposed a new nonlinear Schr\"odiger equation 
$-$ Eq. (\ref{use}) or (\ref{eq1}) $-$ to 
study the properties of a BEC or a general 
superfluid Bose gas valid in both the weak-coupling  and strong-coupling 
(or unitarity) limit. We call this equation the USE (unitary 
Schr\"odiger 
equation).
This equation has a complicated nonlinearity 
structure in its dependence on scattering length $a$ and number of 
atoms $N$. 
In the extreme weak-coupling limit the USE (\ref{use}) reduces to the 
usual 
mean-field GP equation (\ref{gpe}) \cite{rmp}; for medium coupling it 
becomes the  
modified GPE (MGPE) (\ref{mgpe2})  introduced by Fabrocini and Polls 
\cite{polls1}, which is a generalization of the GP equation 
valid for medium coupling incorporating the correction to the 
bulk chemical 
potential of a Bose gas as introduced by Lee, Yang, and Huang 
\cite{huang}. 
However, for very strong coupling, both the GP and 
the MGP equations break down and the bulk chemical potential attains a 
saturation \cite{cowell}. Considering this bulk chemical potential 
Cowell {\it et al.} \cite{cowell} derived a nonlinear equation for a 
superfluid Bose gas  valid in the strong coupling limit. The USE reduces 
to the equation by Cowell {\it et al.} in the strong coupling limit and 
thus have the correct form in both the weak and strong-coupling
regimes. Hence this equation should be useful to study the crossover 
of a superfluid Bose gas from the weak to strong-coupling limits.
In the time-independent form the USE is useful to study the stationary 
properties of a BEC. The full time-dependent USE 
can be used to study  non-equilibrium
properties of a dilute BEC, such as, dynamical 
oscillations \cite{cn}, collapse \cite{col}, free expansion after 
release from the trap \cite{ska3}, 
soliton formation and soliton dynamics \cite{ska,sol} etc.   

In this paper we applied the USE to study vortices in a superfluid Bose 
gas. First, we study vortices in a uniform Bose gas as the scattering 
length is varied from weak to strong coupling values. The vortex core 
radius decreases with increasing scattering length eventually attaining 
a saturation value. The vortex core radius is comparable to the healing 
length in this case. 

Next, we study vortices in a BEC confined  by an axially-symmetric 
harmonic trap by solving the USE numerically. It is found that the 
effective nonlinearity of the USE 
saturates with the increase of scattering length $a$ in the 
strong-coupling 
limit for a fixed number of particles $N$.   In this limit the chemical 
potential as well as the rms sizes of the BCS saturate attain constant 
values. However, the rms sizes obtained from the USE and the MGPE grows 
indefinitely as the scattering length is increased towards   the 
unitarity limit. In the weak-coupling limit the results of the USE are 
compatible with those obtained from the GPE and the MGPE. 
{Finally, we present results of axial and 
radial frequencies of collective 
oscillation of a BEC in a cigar-shaped trap using the USE.} 

The results of this paper for a BEC confined in a axially-symmetric 
harmonic trap can be tested experimentally with present technology 
specially in the unitarity limit near a Feshbach resonance 
\cite{cornish} and this 
will provide a stringent test for the proposed USE.

S.K.A. was partially supported by FAPESP
and CNPq (Brazil), and the Institute for Mathematical Sciences of 
National University of Singapore. Research was (partially) completed 
while S.K.A. was visiting the Institute for Mathematical Sciences, 
National University of Singapore in 2007. 
L.S. was partially supported by GNFM-INdAM and Fondazione CARIPARO 
and thanks Francesco Ancilotto, Arturo Polls,
Luciano Reatto and Grigori Volovik 
for useful discussions.

\end{document}